\documentclass[
aps,prl,twocolumn,
preprintnumbers,superscriptaddress,
]{revtex4-2}

\usepackage[bookmarks=true,colorlinks,linkcolor=blue,urlcolor=cyan,citecolor=red]{hyperref}
\usepackage{xcolor}
\usepackage{amsmath, amssymb, mathtools}

\usepackage{graphicx}
\usepackage{latexsym}
\usepackage{amsfonts}
\usepackage{amssymb}
\usepackage{amsmath}
\usepackage{bm}
\usepackage{mathrsfs}
\usepackage{slashed}
\usepackage{ascmac}
\usepackage{comment}

\usepackage{dcolumn}

\allowdisplaybreaks[3]

\usepackage[
bookmarks=true,colorlinks,linkcolor=blue,urlcolor=cyan,citecolor=red]{hyperref}

\newcommand{\cout}[1]{ \if 0 {#1} \fi }

\newcommand{\beq}{\begin{eqnarray}}
\newcommand{\eeq}{\end{eqnarray}}
\newcommand{\bseq}{\begin{subequations}}
\newcommand{\eseq}{\end{subequations}}
\newcommand{\nn}{\nonumber}
\renewcommand{\=}{&=&}
\newcommand{\nnb}{\nonumber \\}
\newcommand{\pd}{\partial}

\newcommand{\sla}{ \slashed }

\renewcommand{\a}{\alpha}

\newcommand{\ep}{ \epsilon }

\newcommand{\gam}{ \gamma }

\newcommand{\bx}{{\bm{x}}}

\newcommand{\bp}{{\bm{p}}}
\newcommand{\bq}{{\bm{q}}}

\newcommand{\bB}{{\bm B}}
\newcommand{\bE}{{\bm E}}

\newcommand{\M}{ {\mathcal M} }

\newcommand{\prj}{ {\mathcal P} }

\newcommand{\para}{ \parallel}

\newcommand{\tr}{ {\rm tr} }
\newcommand{\diag}{ {\rm diag} }
\newcommand{\sgn}{ {\rm sgn} }

\newcommand{\LLL}{ {\rm LLL} }

\usepackage{color} 
\usepackage[normalem]{ulem}  

\graphicspath{{./figs/}}

\renewcommand{\section}[1]{{\it #1}---}

\newcommand{\ec}{ e }
\newcommand{\qB}{ \xi_\perp }
\newcommand{\Bsp}{ { \mathcal D} }

\newcommand{\Ritus}{ {\mathcal R} }

\begin{document}

\title{ 
Anomaly-Induced Phenomena with Massive Fermions: 
\\
Higher-Landau-Level Dominance from Spatially Modulated Electric Fields
}

\author{Koichi Hattori}
\affiliation{Zhejiang Institute of Modern Physics, Department of Physics, Zhejiang University, Hangzhou, Zhejiang 310027, China}
\affiliation{Research Center for Nuclear Physics, Osaka University, Osaka 567-0047 Japan.}

\author{Kazuya Mameda}
\affiliation{Department of Physics, Tokyo University of Science, Tokyo 162-8601, Japan}
\affiliation{RIKEN iTHEMS, RIKEN, Wako 351-0198, Japan}

\author{Takeru Uchiyama}
\affiliation{Department of Physics, Nagoya University, Furo-cho Chikusa-ku, Nagoya 464-8602 Japan}

\author{Di-Lun Yang}
\affiliation{Institute of Physics, Academia Sinica, Taipei, 11529, Taiwan}
\affiliation{Physics Division, National Center for Theoretical Sciences, Taipei, 106319, Taiwan}

\preprint{RIKEN-iTHEMS-Report-26}

\begin{abstract}

We investigate the axial Ward identity for massive fermions under a constant magnetic field at arbitrary strength, maintaining an arbitrary spacetime configuration of a perturbative electric field.  
We show that a spatially modulated electric field prevents the exact cancellation between the anomaly and pseudoscalar terms, generating a local axial-charge source even in the adiabatic regime where the frequency is subthreshold to massive-fermion production. 
Remarkably, unlike conventional magnetic responses, this charge generation is dominated not by the contribution of the lowest Landau level, but by those of the higher Landau levels. 
Our findings provide a microscopic foundation for anomaly-induced transport and anomalous optical responses in gapped systems. 
In particular, we find that the spatially modulated chiral magnetic effect in weakly gapped Weyl semimetals that exhibits a linear suppression of the magneto-resistance by the magnetic-field strength instead of the renowned quadratic suppression. 
\end{abstract}

\maketitle

\section{Introduction}%
Chiral anomaly is a prominent facet of quantum fluctuations in electromagnetic fields, rendering chiral symmetry broken and thus the axial current nonconserved at the quantum level~\cite{Adler:1969gk, Bell:1969ts}. 
Axial charge, or chirality imbalance, is produced via chiral anomaly and induces novel phenomena both in high-energy and condensed-matter systems, including relativistic heavy-ion collisions, cosmology/astrophysics, and Dirac/Weyl semimetals (see Refs.~\cite{Miransky:2015ava, Hattori:2016emy, Landsteiner:2016led, Armitage:2017cjs, Hidaka:2022dmn, Kamada:2022nyt, felser2023topology, Hattori:2023egw} for recent reviews). 
The axial Ward identity (AWI) governs axial-charge production and serves as a key equation for anomaly-induced phenomena. 

Although the theoretical framework for massless fermions is well established as an idealized limit~\cite{Miransky:2015ava, Hattori:2016emy, Landsteiner:2016led,Hidaka:2022dmn,  Kamada:2022nyt, Hattori:2023egw}, 
its application to realistic massive fermions faces a fundamental barrier 
since explicit chiral symmetry breaking by a nonzero fermion mass gives rise to relaxation of the axial charge. 
This relaxation effect is captured by the pseudoscalar (PS) term in the AWI and competes with the axial-charge pumped by the anomaly term. 
In the limit of a constant electromagnetic field, this PS term exactly cancels the anomaly term~\cite{Schwinger:1951nm, Ambjorn:1983hp, Copinger:2018ftr}, completely washing out the axial-charge source and precluding experimental realization of anomaly-induced phenomena.

In this Letter, however, we show that this complete washout is a property of the constant electromagnetic field limit, and thus can be  
prevented by a finite momentum scale competing with a fermion mass. 
To apply this mechanism to magnetic-field-induced anomalous phenomena, we investigate the AWI in magnetized Dirac systems across all Landau levels, and reveal that the interplay between a constant magnetic field and a spatially modulated electric field gives rise to a local axial-charge source.

This new mechanism overrides our understanding of the axial-charge production in the following two aspects. 
First, the axial charge is generated even in the {\it adiabatic regime} i.e., even when the frequency of the electric field is much smaller than the fermion mass and real-particle production is kinematically forbidden. 
This contrasts with the previously studied mechanisms via real-particle production by the Schwinger mechanism in a nonperturbative electric field~\cite{Copinger:2018ftr} or by sufficiently high-frequency electric fields \cite{Hattori:2022wao}.  
Second, the produced axial charge 
is dominated by the contributions from the higher Landau levels (hLLs), which we call the {\it higher-Landau-level dominance} in the axial-charge production by a spatially modulated electric field. 
This also fundamentally contrasts with the conventional argument based on the lowest-Landau-level (LLL) approximation~\cite{Nielsen:1983rb, Fukushima:2008xe,Landsteiner:2013sja,Landsteiner:2016led,Armitage:2017cjs, Ong:2020ffe,Hattori:2022wao}, readily implying that the saturation of the chiral anomaly~\cite{Adler:1969er, Itoyama:1982up} is not ``LLL exact'' in a spatially modulated electric field but involves contributions from all the Landau levels. 

To demonstrate physical implications of our theoretical findings, we evaluate the anomalous photon dispersion relations and the electric conductivity, that is, the spatially modulated chiral magnetic effect (CME). 
We find the anomaly-induced birefringence including the mass effect 
and the linear suppression of the magneto-resistance by the magnetic-field strength $\sim 1/|\bB|$, instead of the renowned quadratic suppression~\cite{Son:2012bg, Li:2014bha}. 

We use the mostly minus sign convention for the Minkowski metric $g^{\mu\nu}=g^{\mu\nu}_\perp+g^{\mu\nu}_\parallel$ with $g^{\mu\nu}_\perp = \diag(0, -1 ,-1, 0)$ and $g^{\mu\nu}_\parallel = \diag(1, 0, 0,-1)$, and define $v^\mu_\perp = g^{\mu\nu}_\perp v_\nu$ and $v^\mu_\parallel = g^{\mu\nu}_\parallel v_\nu$ for a vector $v^\mu$.

\begin{figure}
     \begin{center}
              \includegraphics[width=0.9\hsize]{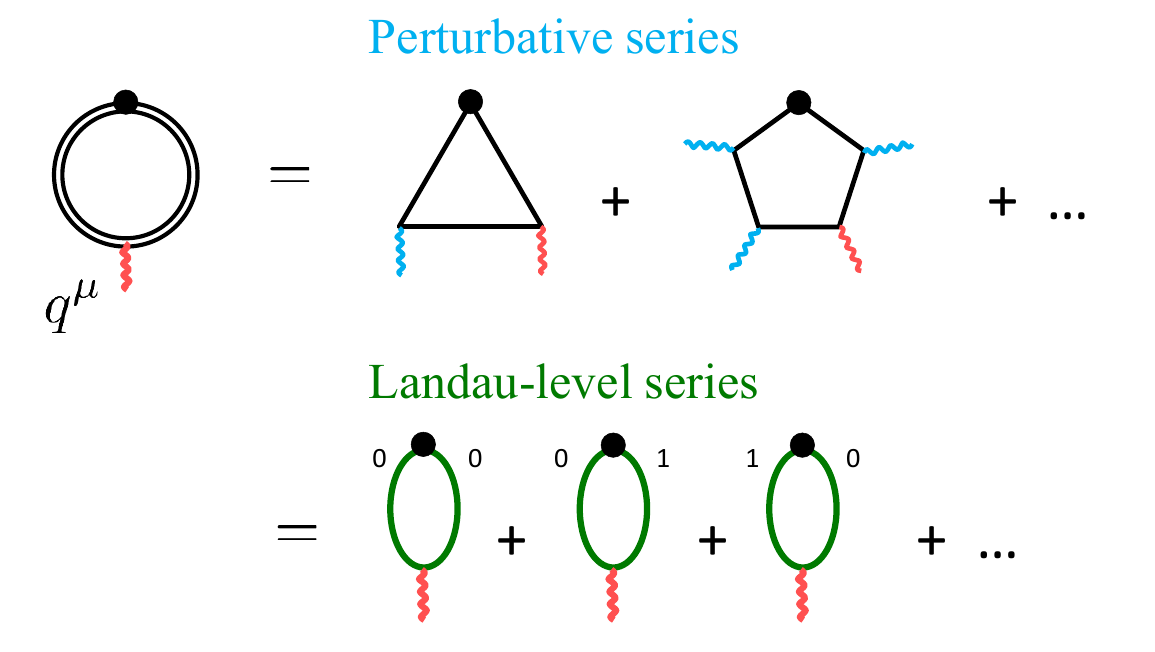}
     \end{center}
\vspace{-0.5cm}
\caption{Axial current in response to a perturbative electric field (wavy red lines) in the presence of a constant magnetic field (blue wavy lines), expanded in a perturbative series and in a Landau-level series. 
The triangle anomaly is shared by all the Landau levels. 
}
\label{fig:triangle2}
\end{figure}

\section{Anomaly diagrams in the Landau levels}We construct the AWI in a constant magnetic field $\bB=(0,0,B)$ at arbitrary strength and a perturbatively weak electric field $\tilde \bE(q)$ at arbitrary frequency and spatial momentum, $q^\mu$. 
This construction demands a resummation of the Feynman diagrams with perturbative insertion of the magnetic field in Fig.~\ref{fig:triangle2}, 
where the blue and red wavy lines show the constant magnetic field and the perturbative photon field, respectively. 
Resummation of the perturbative series with respect to the magnetic field can be reorganized into another series with respect to the Landau levels 
that are encoded in the resummed fermion propagator (\ref{eq:resummed-prop}). 
These are two different series representations of the same AWI. 

The matrix element for the divergence of the axial current reads $\int d^4 x \, e^{-iq\cdot x} \langle k |\pd _\mu j_{\rm A}^\mu (x) | 0\rangle
= (2\pi)^4 \delta^{(4)}( k -q) i q_\mu  \M^{\mu\nu}(q)  \epsilon^\ast_\nu (k)$.  
The momentum is conserved between the axial current and the photon field since an external constant magnetic field has a vanishing momentum. 
Coupling to the photon field, with a polarization vector $ \epsilon_\nu (k) $, will be finally given by the Fourier spectrum of the electric field $ \tilde E^i (k) =  i ( k^0 A^i - k^i A^0)$. 
The above matrix element is written down as 
\begin{eqnarray}
i \M^{\mu\nu} \=  - \ec \int \frac{d^4p}{(2\pi)^4} \tr\big[ \gamma^{\mu} \gam^5 S(p ) \gam^\nu S(p+q ) \big ] .
\label{eq:Matrix}
\end{eqnarray}
Only one difference from the standard Feynman rule 
\cite{Peskin:1995ev} is that the fermion propagator $  S(p )$ 
is replaced by a resummed propagator in a magnetic field 
\cite{Schwinger:1951nm, Chodos:1990vv, Gusynin:1995nb, Hattori:2023egw}: 
\begin{eqnarray}
 S(p) = \sum_{n=0}^\infty 
\frac{ 2 i (-1)^n \Bsp_n (\frac{|\bp_\perp|^2}{2|eB|}) }{ p_\parallel^2 - m^2   - 2n \vert \ec B \vert + i\epsilon  }
e^{ -\frac{|\bp_\perp|^2}{2|eB|}} . \label{eq:resummed-prop}
\end{eqnarray} 
The pole positions are specified by 
the longitudinal momentum $p_\para^2 $
as a manifestation of the Landau quantization. 
The Dirac-spinor structure is given as 
$ \Bsp_n (x) =
( \slashed p _\parallel + m ) \big\{ 
\prj_+ L_n(x) - \prj_-  L_{n-1} (x) \big\}
- 2 \slashed p _\perp L_{n-1}^1 (x)$
with the associated Laguerre polynomial $ L_n^\a(x) $ and $L_{n}(x)\coloneqq L^{0}_{n}(x)$ and the spin projection operator $\prj_\pm = \frac12 [1\pm i \, \sgn(\ec B) \gam^1 \gam^2]$. 
The appearance of these factors reflects 
the fermion wave function in the Landau levels and the Zeemann effect. 
There is no state with a negative index, so we define $ L_{-1} (x)\coloneqq 0 $ and $ L^1_{-1} (x)\coloneqq0$. 

One can proceed by performing the Dirac trace, introducing the Feynman parameter, and carrying out the momentum integral using the standard computational procedure. 
Then, we arrive at a simpler expression 
\begin{eqnarray}
  i q_\mu \M^{\mu\nu} \= - i \frac{\ec^2 B}{2 \pi^2} \epsilon^{\nu\a}_\parallel q_{\parallel \a} w\big( \lambda , \qB,b \big) ,
\end{eqnarray}
with $\epsilon_\parallel^{\mu\nu}$ being the Levi-Civita symbol for the longitudinal components, $\epsilon_\para^{03} = 1$. 
Equivalently, the AWI in momentum space is written as
\begin{eqnarray}
\label{eq:AWI}
iq_{\mu} \langle k|\tilde{j}_\mathrm{A}^{\mu}(q)|0\rangle =\frac{e^2}{2\pi^2}\langle k|\tilde{\bm E}(q)|0\rangle\cdot {\bm B} \,w\big( \lambda , \qB,b \big).
\end{eqnarray}
Here, $w$ is a function of the three dimensionless parameters: $\xi_\perp \coloneqq  |\bq_\perp|^2/(2|eB|)$, $\lambda \coloneqq q_\para^2/m^2$, and $b\coloneqq2|eB|/m^2$. 
This weight function is decomposed as
$w= w_{\rm anom} + w_{\rm PS}$; these two correspond to the contributions from chiral anomaly and the PS term, respectively, as we will see below. 
In the former term, we assemble all the mass-independent terms
\begin{eqnarray} 
\label{eq:w_a}
w_{\rm anom}
\coloneqq \sum_{n,n'=0}^\infty 
(I_{n,n'} - I_{n-1,n'-1}) ,
\end{eqnarray} 
where
\begin{eqnarray} 
I_{n,n'} (\qB)
 \=  (-1)^{n+n'} \frac{8\pi}{ |e B|} 
 \int \frac{d^2 p_\perp}{(2\pi)^2} e^{ -\frac{ |\bp_\perp|^2  }{|eB |} }
  L_n \Big(2 \frac{ |\bp_\perp|^2 }{|eB |}\Big)  
\nnb
 &&\times
 e^{ -\frac{  |\bp_\perp + \bq_\perp|^2 }{|eB |} }
 L_{n'} \Big(2 \frac{ |\bp_\perp + \bq_\perp|^2 }{|eB |} \Big)
 \nnb
 \= \frac{ n!}{ n'! } \,  e^{ -\qB} \, \qB^{n'-n} \left[ L_n^{n'-n} ( \qB) \right]^2.
 \label{eq:I_nn}
\end{eqnarray}
This $I_{n,n'}$ stems from the transverse-momentum integral with convolution of the fermion wave functions~\cite{Fukushima:2024ete}.
The two terms $I_{n,n'}$ and $I_{n-1,n'-1}$ correspond to the contributions from the two-fold spin-degenerate hLLs, 
while the LLL contribution is given by 
the single term $I_{0,0}=e^{ -\qB}$. 
The other mass-dependent term $w_{\rm PS}$ is expressed as
\begin{eqnarray}
 w_{\rm PS} \coloneqq 
- \sum_{n,n'=0}^\infty 
(I_{n,n'} - I_{n-1,n'-1})  \int_0^1  \frac{m^2 \, dx}{ \Delta_{n,n'} -i\epsilon} 
,
\label{eq:w_PS}
\end{eqnarray}
where $ \Delta_{n,n'}(x) \coloneqq  x m_n^2 + (1-x) m_{n'}^2 - x(1-x) q_\para^2$ with the energy gap $m_n^2 = m^2 + 2n |\ec B| $ in the Landau levels.

The telescoping series $w_\mathrm{anom}$ can be evaluated with the generating function of $I_{n,n'}$, i.e., 
\begin{equation} 
\label{eq:Gen}
 G(t,s) \coloneqq \sum_{n,n'=0}^\infty t^n s^{n'} I_{n,n'}
  =  \frac{e^{-\frac{(1-t)(1-s)}{1-ts}\xi_\perp}}{1-ts} ,
\end{equation}
where $|t|<1, \,|s|<1 $, and the rightmost side is obtained from the generating function of the Laguerre polynomial (see Appendix). 
Namely, the series $w_\mathrm{anom}$ in Eq.~(\ref{eq:w_a}) is expressed with $ G(t,s)$ in the limit $t,s\to 1$, leading to a simple result
\begin{equation}
\label{eq:K=1}
\begin{split}
 w_{\rm anom}
 &= 
 \lim_{t,s\to 1}(1-ts)G(t,s)
=1 \, .
\end{split}
\end{equation}   
It is found that summing over all the Landau levels gives a constant whereas each of $I_{n,n'}$ depends on a specific value of $\xi_\perp$. 
Inserting $w_{\rm anom}=1$ into Eq.~(\ref{eq:AWI}) shows that the summation over all the Landau levels is required to reproduce chiral anomaly in (3+1) dimensions. 
In other words, chiral anomaly is, in general, not saturated by 
the LLL contribution
\begin{equation}
\label{eq:anom_LLL}
 w_{\rm anom}^\LLL = I_{0,0} = e^{-\xi_\perp}
 ,
\end{equation}
although the LLL has the (1+1)-dimensional linear dispersion~\cite{Nielsen:1983rb}.

\section{Local axial-charge source in the adiabatic limit}%
In contrast to the universal anomalous term, $w_\mathrm{anom}=1$, the PS term $w_{\rm PS}$ depends on the parameters $\xi_\perp$, $\lambda$, and $b$, defined below Eq.~(\ref{eq:AWI}). 
To warm up, we discuss the conventional limits as a baseline. 
We define the massless limit as $|\lambda|\to \infty$ at arbitrary $B$. 
In this limit, one easily finds that $w_{\rm PS} =0$, and the AWI is solely governed by chiral anomaly, i.e., $w = w_{\rm anom}=1$. 
In the opposite extreme, we also define the adiabatic limit as $\lambda\to 0$, 
which has three cases of the hierarchy examined below. 
First, when we consider $|q_\para^2| \ll |\ec B|\ll m^2$, 
we have $ \Delta_{n,n'} \to  m^2$, leading to complete cancellation between the anomalous and PS terms in each Landau level, i.e., $w=0$. 
This is analogous to the usual washout of the axial charge, which is interpreted with the spectral flows on quadratic dispersion relations~\cite{Ambjorn:1983hp} (see also Ref.~\cite{Landsteiner:2016led}). 

In another hierarchy of the adiabatic limit, $|q_\para^2| \ll m^2 \ll |\ec B|$, however, we show that $w\neq 0$. 
Since $\Delta_{0,0}(x) \to m^2$, but $\Delta_{n,n'}(x) \to |\ec B|$ for all the other pairs of $(n,\,n')$, a nonzero contribution to the PS term only comes from the LLL, leading to 
\begin{eqnarray}
\lim_{b\to \infty} 
\, \lim_{\lambda\to 0} \, 
w_\mathrm{PS} = - e^{- \qB}  \, .
\label{eq:w-q-m-B}
\end{eqnarray} 
While the LLL contribution to the anomaly term~\eqref{eq:anom_LLL} is exactly compensated by this PS term~\eqref{eq:w-q-m-B}, the hLL contribution $w_\mathrm{anom}^\mathrm{hLL}=w_\mathrm{anom}-w_\mathrm{anom}^\mathrm{LLL}=1-e^{-\xi_\perp}$ is not.
Consequently, the AWI acquires a nonzero source term proportional to $w = 1 - e^{-\xi_\perp}$ even for gapped fermions and in the adiabatic limit. 
Note that the suppression of hLL contributions to the PS term is understood from the fact that the hLLs are approximate chirality eigenstates when $m^2 \ll |\ec B|$, and the chirality mixing is suppressed accordingly, in spite of their quadratic dispersion relations; 
we elaborate on this point in the Appendix. 

This generation of an axial-charge source in inhomogeneous fields even for gapped fermions in the adiabatic limit is our main result in this Letter, in contrast to the complete washout in homogeneous fields recapitulated above.
A remarkable feature of this generated charge is its local nature, distinct from a global chirality imbalance.
Indeed, the spatial-volume integral of the AWI picks up the homogeneous Fourier component $\bq_\perp=0$ (i.e., $\xi_\perp=0$), which yields $w=0$ and reproduces the usual washout of the global axial charge~\cite{Schwinger:1951nm, Ambjorn:1983hp, Copinger:2018ftr}.
Rather than a global accumulation, the charge is locally generated over the magnetic length $1/\sqrt{|eB|}$, which is a typical scale for the spatial extension of the wave function. 
Consequently, this finite local source can be probed by a spatially modulated electric field with a finite transverse momentum $\xi_\perp \gtrsim 1$.

For completeness, we also discuss the weak-field limit in the Appendix 
that covers the third case of the adiabatic limit, $|\ec B|\ll  |q_\para^2| \ll m^2$. 
In fact, this is the most technically challenging limit since 
one cannot simply take this limit in the Landau-level series. 
We confirm that the isotropy in the momentum space is correctly recovered.

\section{Applications}%
Let us discuss applications of the local axial-charge source to the anomaly-induced phenomena. 
We focus on the interplay between modulated electric fields and ultra-relativistic fermions with the scale separation such that the typical momentum of quasi-particles $p$ is much greater than the mass gap and the four-momentum of the electric field, $p\gg m\sim q$.

We first show that the dispersion relation of photons traversing Weyl semimetals acquires a polarization-dependent modification, i.e., birefringence. 
Let us consider Maxwell's equations in momentum space, 
\begin{eqnarray}\label{eq:Maxwell's_eqs}
	i{\bm q}\times\tilde{\bm E}=i\omega\tilde{\bm B},\quad -i{\bm q}\times\tilde{\bm B}=i\omega \tilde{\bm E}+\tilde{\bm j}, 
\end{eqnarray}
with $q^{0}=\omega$, where the electromagnetic fields and charge current stem from the fluctuations,
$\tilde{\bm E}=\delta \tilde{\bm E},\quad \tilde{\bm B}={\bm B}_{\rm ex}+\delta\tilde{\bm B},\quad \tilde{\bm j}=\delta\tilde{\bm j}$, 
under a constant background magnetic field, ${\bm B}_{\rm ex}\gg\delta{\bm B}$. Hereafter we use $\delta O$ to represent the fluctuation of an arbitrary physical quantity $O$. Our goal is to obtain the dispersion relation of $\delta{\bm E}$ from Maxwell's equations and fluctuating currents in connection with the AWI.

We may write down the charge current linear to the fluctuations as 
$\delta \tilde{\bm j}=\sigma \delta\tilde{\bm E}+\delta\xi_{B}  {\bm B}_{\rm ex} 
$, for the coupling $\ec$ absorbed into fields hereafter, when the ``effective" CME conductivity $\delta\xi_{B}$ is generated by the electromagnetic dynamics via the AWI. 
Here $\sigma$ is the usual Ohmic conductivity depending on the interactions, and we treat it as a phenomenological constant for simplicity. 
On the other hand, $\delta\tilde{\bm j}_\mathrm{A} $ may be triggered by the chiral separation effect (CSE), $\delta\tilde{\bm j}_\mathrm{A} =\sigma_{5}\delta\tilde{\bm B}=\sigma_{5}{\bm q}\times \delta\tilde{\bm E}/\omega$, with Eq.~(\ref{eq:Maxwell's_eqs}) used to reach the second equality.
Following the setup in Weyl semimetals like Refs.~\cite{Li:2014bha,Gorbar:2021ebc}, we may introduce the CME and CSE conductivity as $\xi_{B}= \mu_5/(2\pi^2)$ and $\sigma_{5}= \mu_{V}/(2\pi^2)$, where $\mu_V$ denotes the vector chemical potential
and the axial chemical potential $\mu_5$ is related to an axial-charge density in thermal equilibrium, $n_5\approx \mu_5\chi$ with $\chi(T,\mu_V) =(T^2+3\mu_{V}^2/\pi^2)/(3v_F^3)$ for $T,\mu_V\gg \mu_5$ and $v_F$ being the Fermi velocity. Now, we augment the AWI with an additional relaxation term due to the inter-cone transition, 
\begin{eqnarray}
\label{eq:anomaly_tau}
	\omega \delta\tilde{n}_{5} 
	\=- \frac{ i\delta\tilde \bE(q) \cdot \bB_{\rm ex}}{2\pi^2 }     w( \lambda , \qB, b )+\frac{i\delta \tilde{n}_{5}}{\tau_{5}},
\end{eqnarray}
where $\tau_{5}$ represents a constant axial relaxation time and we have taken ${\bm q}\cdot\delta\tilde{\bm j}_{A}=0$ from $\delta\tilde{\bm j}_{A}\propto{\bm q}\times \delta\tilde{\bm E}$. 
Thus, the CME conductivity is expressed with $w( \lambda, \qB ,b)$ as 
\begin{eqnarray}\label{eq:CME_conductivity}
\delta\xi_{B}=\frac{  \tau_5\delta\tilde \bE(q) \cdot \bB_{\rm ex} w( \lambda, \qB ,b)}{(2\pi^2)^2 \chi\left(1+i\tau_5\omega\right)}  .
\end{eqnarray}

For the photons polarized along the magnetic field with $\delta\tilde{\bm E}\parallel {\bm B}_{\rm ex}$ and ${\bm q}\cdot\delta\tilde{\bm E}=0 \ (q_z=0)$, 
combining Eq.~(\ref{eq:Maxwell's_eqs}) and the explicit expression of $\delta\tilde{\bm j}$ results in the dispersion relation 
\begin{eqnarray}\label{eq:dis_Weyl_semimetals}
	\omega^2=|\bm q|^2+i\omega\sigma+\frac{  |\bB_{\rm ex}|^2w( \lambda, \qB ,b)}{(2\pi^2)^2 \chi \left[1-i(\omega\tau_5)^{-1}\right]}    .
\end{eqnarray}
We then analyze the case with $\sigma\rightarrow 0$ and $(\omega\tau_5)^{-1}\rightarrow 0$. For low-frequency photons such that $\omega\ll m$ in Eq.~(\ref{eq:w-q-m-B}), one can approximate $w\approx 1-e^{-\qB}\approx |{\bm q}|^2/|2{\bm B}_{\rm ex}|$ and derive 
the phase velocity
\begin{eqnarray}
v_{p}\coloneqq\frac{\omega}{|{\bm q}|}=\left(1+\frac{  |\bB_{\rm ex}|}{8\pi^4 \chi }\right)^{1/2}.
\end{eqnarray} 
Conversely, in the opposite limit for $\omega\gg m$ and accordingly $w\approx 1$, the phase velocity becomes 
\begin{eqnarray}
	v_{p}=\left(1+\frac{ |\bB_{\rm ex}|^2}{8\pi^4 |{\bm q}|^2\chi}\right)^{1/2}.
\end{eqnarray}  
See Ref.~\cite{Hanai:2025pkw} for a similar study in such a limit.
Recall that the other polarization mode with $\delta\tilde{\bm E}$ perpendicular to ${\bm B}_{\rm ex}$ remains light-like and $v_{p}=1$. 
We thus found the anomaly-induced birefringence including the mass effect.

Moreover, Eq.~(\ref{eq:CME_conductivity}) has another important application. 
Taking $\omega=|{\bm q}|=0$ for constant electric fields in Eq.~(\ref{eq:CME_conductivity}), one finds that $w=0$ and thus the CME current vanishes, as expected.
However, we may apply a static electric field parallel to the background magnetic field ${\bm B}_{\rm ex} = |\bB_{\rm ex} |\hat{\bm{z}}$, but with its amplitude spatially modulated in the transverse plane with a finite ${\bm q}_\perp$, whereas $\omega =  q_z = 0$ (thus $q_\para^\mu = 0$).  
By substituting the static limit of $w(0,\qB,b)$ from Eq.~(\ref{eq:w-q-m-B}), 
it is found that 
\begin{eqnarray}\label{eq:xiB}
	\delta\xi_{B} =\frac{\tau_5\delta\tilde \bE(q) \cdot \bB_{\rm ex} (1 - e^{-\frac{|{\bm q}_{\perp}|^2}{2|{\bm B}_{\rm ex}|}})}{(2\pi^2)^2 \chi}
    .
\end{eqnarray}
This spatially modulated CME is a consequence of the local axial-charge source with the hLL dominance. 

The spatially modulated current density gives rise to a current $\mathcal{J}(R)
\coloneqq\int^{R}_{0} dr\int^{2\pi}_{0}d\theta \, r\, \delta j^{z}(\bx_\perp)$ 
flowing through a finite-volume cylinder along its axis per unit length, 
where $\bx_\perp$ is the transverse coordinate.
Here, parallel magnetic and electric fields are applied. 
Now, taking $\delta\bE(\bx_\perp)=E_{0}e^{-r^2/R_{E}^2}\hat{\bm z}$ as a concrete example, one can show that 
\begin{eqnarray}
	\delta j^z(\bx_\perp)=\frac{\tau_5|\bB_{\rm ex}|^2E_0}{ 4\pi^4 \chi}
    e^{-\frac{r^2}{R_E^2}}
    \left(1-\frac{e^{\frac{2r^2}{R_E^2(2+\bar{B})}}}{ 1+ 2/ \bar{B} }\right),
    \label{eq:current-density-result}
\end{eqnarray}	
where $\bar{B}\coloneqq|{\bm B}_{\rm ex}|R_{E}^2$. When $\bar{B}\rightarrow \infty$, corresponding to $|{\bm q}_{\perp}|^2\ll|{\bm B}_{\rm ex}|$, the current reads 
\begin{eqnarray}
    \mathcal{J}(R) 
	=\frac{ \tau_5|\bB_{\rm ex}|E_{0}}{2\pi^3\chi}
    e^{-\frac{R^2}{R_E^2}}
    \left(\frac{R}{R_{E}}\right)^2
    \, ,
\end{eqnarray} 
which reaches a maximum at $R=R_E$.
For $\bar{B}\rightarrow 0$, corresponding to $|{\bm q}_{\perp}|^2\gg|{\bm B}_{\rm ex}|$, it turns out that
\begin{eqnarray}
	 \mathcal{J}(R)  
	=\frac{ \tau_5|\bB_{\rm ex}|^2E_{0}R_E^2}{4\pi^3\chi}\left(1-e^{-\frac{R^2}{R_{E}^2}}\right),
\end{eqnarray} 
which saturates to a maximum when $R/R_E\rightarrow\infty$.
Note that $\xi_{B}\sim \tilde \bE(q) \cdot \bB_{\rm ex}$ and the PS term vanishes according to Eq.~(\ref{eq:xiB}) in the large-$|{\bm q}_{\perp}|^2$ limit. 
In summary, we found the distinct dependence of the current $\mathcal{J}(R)$ on the magnetic-field strength in the regimes specified by the ratio $|{\bm q}_{\perp}|^{2}/|{\bm B}_{\rm ex}|$. 
For weak modulation, the magneto-resistance, which is the inverse of electric conductivity, is proportional to $1/|\bB_{\rm ex}|$, instead of the quadratic dependence $1/|\bB_{\rm ex}|^2$ that has been regarded as a potential signal of CME in a uniform electric field. However, for strong modulation, a similar $1/|\bB_{\rm ex}|^2$ dependence is found.

\section{Summary and outlook}%
In this Letter, we showed that the AWI under a constant magnetic field and a spatially modulated electric field leads to the local axial-charge source that exists even in the adiabatic limit and is dominated by the hLL contributions, overriding our conventional understanding. 
These results with spatially modulated electric fields provide new insight into the study of anomaly-induced phenomena in gapped fermion systems. 
We leave the investigation of the non-adiabatic regimes, where real-particle production can occur, and finite-temperature/radiative corrections to the PS terms for future works. 
Further, it is important to evaluate the conductivity with the Kubo formula and kinetic theory.


\begin{acknowledgments}
\section{Acknowledgments}%
The authors thank Kenji Fukushima, 
Yoshimasa Hidaka, Xu-Guang Huang, Hidetoshi Taya, and Sota Hanai for discussions. 
This work was supported by JSPS KAKENHI Grant Numbers~20K03948, 24K17052, and 23K22487, by National Science and Technology Council (Taiwan) under Grant No. NSTC 113-2628-M-001-009-MY4, by Academia Sinica under Project No.~AS-CDA-114-M01, by National Science Foundation of China under grant No.~W2532002, and by JST SPRING, Grant Number JPMJSP2125. 
\end{acknowledgments}

\vspace{0.5cm}

\appendix

\section{Generating function of the transverse integral}
We derive the generating function of the transverse integral $I_{n,n'}(\xi_\perp)$ defined in Eq.~(\ref{eq:I_nn}).  
To begin, we introduce a function 
\begin{eqnarray} 
G(t,s)  &\coloneqq& 
\sum_{n=0}^\infty \sum_{n'=0}^\infty 
t^n s^{n'} I_{n,n'} \, ,
\label{eq:Gen-def}
\end{eqnarray} 
which generates the integral as 
\begin{eqnarray}
I_{n,n'} =  
\lim_{s,t \to 0} \frac{1}{n! n'!} 
\frac{\pd^n \ }{\pd t^n} \frac{\pd^{n'} \ }{\pd s^{n'}} 
G(t,s) \, .
\label{eq:generating-I}
\end{eqnarray}
According to the definition (\ref{eq:Gen-def}), the explicit form of the generating function is given as 
\begin{eqnarray} 
G(t,s)  &\coloneqq& 
\sum_{n=0}^\infty \sum_{n'=0}^\infty 
(-t)^n (-s)^{n'}  e^{ -\xi_\perp } 
\int \frac{d^2 p_\perp}{\pi} e^{ - | \bp_\perp |^2 } 
\nnb
&& \times
L_n \Big(|\bp_\perp - \sqrt{\xi_\perp} \hat \bq_\perp |^2\Big)  
L_{n'} \Big( |\bp_\perp +\sqrt{\xi_\perp} \hat \bq_\perp  |^2\Big) \, ,
\nnb
\label{eq:Gen-integral}
\end{eqnarray} 
where $ \hat \bq_\perp \coloneqq  \bq_\perp/ |\bq_\perp| $ and the integral variable was rescaled. 
One can reduce the above integral to the Gaussian integral by using the generating function of the Laguerre polynomial 
\begin{eqnarray}
\sum_{n=0}^\infty t^n L_n(x) 
= \frac{1}{1-t} e^{ - \frac{t}{1-t} x} \, .
\end{eqnarray}
Then, it is straightforward to obtain the compact form
\begin{eqnarray}
G(t,s) 
= \frac{ 1 }{ 1- st}  
e^{ -\frac{ (1  -t)(1- s) }{1-ts} \xi_\perp   } \, ,
\end{eqnarray} 
which is shown in Eq.~(\ref{eq:Gen}).

\section{Suppression of chirality mixing in the hLLs}
To understand the suppression of the hLL contributions to the PS term in Eq.~(\ref{eq:w-q-m-B}), 
it is useful to examine the mass-dependent part of the AWI more explicitly. 
As in the perturbative electromagnetic fields discussed in standard textbooks (see, e.g., Ref.~\cite{Peskin:1995ev}), 
after separating the anomalous term from the residual mass-dependent term, the latter can be identified with the matrix element of the PS bilinear $2im\langle \bar\psi\gamma^5\psi\rangle$ in the presence of the external vector perturbation. 
This matrix element quantifies the magnitude of chirality mixing induced by a finite fermion mass.

The mass-dependent terms in the matrix element (\ref{eq:Matrix}) are summarized as 
\begin{eqnarray}
i q_\mu \M^{\mu\nu}_{\rm mass} 
\= 
2m \sum_{n,n'=0}^\infty \sum_{\sigma=\pm} \int_0^1 dx \int \frac{d^2 p_\para}{(2\pi)^2} 
\\
 && \times
 \frac{ \tr[  \gam^5 \prj_\sigma (\sla p_\para +m) \gam_\para^\nu   (\sla p_\para + \sla q_\para+m) ] }
{ [\, \{ p_\para +  (1-x) q_\para \}^2  - \Delta_{n,n'} \, ]^2 }  
\nn
\, .
\end{eqnarray} 
This is a two-point function with $ \gam^5 $ on one of the vertices, and is readily identified with $2im\langle \bar \psi \gam^5 \psi \rangle$. 
It is crucial to note that the spinor trace is nonzero only when one picks up one mass term due to the presence of $\gam^5$. That is, the matrix element is nonzero only when chirality is flipped (odd times) by the mass term on a fermion loop so that chirality flow on the fermion loop is consistent with chirality components of the PS bilinear. 
Then, the fermion-momentum scale is irrelevant in the numerator because the linear momentum term does not contribute to the integral (after the integral variable is shifted by the external photon momentum $q_\para^\mu$).

On the other hand, the denominator suppresses the matrix element due to the energy gap in the dispersion relations, so that the magnitude of the chirality mixing is governed by a ratio of fermion mass to an energy gap. 
Chirality mixing in the hLL is suppressed by a dimensionless parameter $m^2/|eB| \ll 1$, and this is the reason why the hLL contribution to the PS term is suppressed as shown in Eq.~(\ref{eq:w-q-m-B}).

One can also show the suppression of chirality mixing for the hLL in the strong magnetic-field limit, $|eB| \gg m^2$, 
by explicitly solving the Dirac equation 
\begin{eqnarray}
\left( i \slashed D - m \right) \psi = 0
\label{eq:Dirac}
\, ,
\end{eqnarray}
where the covariant derivative in an external field is defined as 
$ D^\mu = \partial ^\mu + i e A^\mu(x)$ with $A_0=A_3=0$.  
In the following discussion, we only use a gauge-invariant algebra 
$ [ D^1 , D^2] 
= - i e B$. 
An upshot of this discussion is that the hLL has a large transverse momentum of the order of $\sqrt{|eB|}$ associated with the cyclotron motion, and thus the chirality mixing is suppressed in a similar manner to free relativistic particles.

An explicit solution for this Dirac equation can be written as 
\begin{eqnarray}
\psi (x) =   e^{ - i p_\para^\mu x_\mu} 
\Ritus_{n} (x_\perp)  \, u
\label{eq:Dirac-Ritus}
\, ,
\end{eqnarray}
with the Ritus basis $ \Ritus_{n} (x_\perp) $ 
\cite{Ritus:1972ky, Ritus:1978cj} 
and a four-component spinor $ u$ determined below. 
The Ritus basis is given as 
\begin{eqnarray} 
\Ritus_{n} (x_\perp) 
=  \phi_{n} (x_\perp) \prj_+ +  \phi_{n-1} (x_\perp) \prj_- 
\label{eq:Ritus}
\, ,
\end{eqnarray}
where $  \phi_{n} $ is the $ n$-th energy eigenfunction 
of the Klein-Gordon equation $( D_\mu D^\mu  + m^2) \phi =0$ 
and $  \phi_{-1} \coloneqq 0 $ is promised. 
The two terms on the right-hand side correspond to 
the twofold-degenerate spin states.

Applying the Dirac operator to the Ritus basis (\ref{eq:Dirac-Ritus}), 
one can show that (see, e.g., Ref.~\cite{Hattori:2023egw})
\begin{eqnarray}
i \sla D  \, \psi(x) 
= e^{ - i p_\para^\mu x_\mu}   \Ritus_{n} (x_\perp)  
\big( \sla p _\para - \sqrt{2n |eB| }\ \gam^1   \big) \, u
\, .\ \label{Ritus_u}
\end{eqnarray}
Therefore, the ansatz (\ref{eq:Dirac-Ritus}) solves the Dirac equation (\ref{eq:Dirac}) 
if the spinor $ u $ solves a ``free'' Dirac equation 
\begin{eqnarray}
 ( {\sla p}_n -m ) u ( p_n) = 0
 \label{eq:free}
 \, ,
\end{eqnarray}
with a four-momentum 
$ p^\mu_n  \coloneqq ( \ep_n , \sqrt{2n |eB| },0,p^3) $ 
and the Landau levels $ \ep_n = \sqrt{p_z^2 + 2n |eB|  + m^2}$. 
Because of a commutative property $[\gam^5, \prj_\pm] = 0$, $\psi$ in Eq.~(\ref{eq:Dirac-Ritus}) is a chirality eigenstate if $u(p_n)$ is. 
In a massive case, the chirality-mixing magnitude is strongest near the bottom of the parabolic dispersion relations where $ p_z \lesssim m$. 
This is always the case for the LLL $(n=0) $ since $ p^\mu_0 = ( \ep_0 , 0,0,p^3) $ is independent of $|eB|$. 
However, for the hLL $ (n\geq 1)$, 
the mixing effect in $u(p_n)$ is suppressed in a strong magnetic field because the magnitude of the momentum is bounded as $|\bp_n|^2 \geq 2n|eB| \gg m^2$. 
Similar to particles in free space, $\psi$ approaches a chirality eigenstate as $m^2/|\bp_n|^2 \leq m^2/|eB| \to 0 $ even though the hLL dispersion relations have the quadratic forms.

\section{Derivation of the local current} 
For the inhomogeneous fluctuation of an electric field in momentum space, we may factorize it as $\delta\tilde \bE(q)=(2\pi)^2\delta^2(q_{\parallel})\delta\tilde \bE({\bm q}_{\perp})$ and evaluate the inverse Fourier transform
\begin{eqnarray}
&&\hspace{-0.5cm}
\int \frac{d^2{\bm q}_{\perp}}{(2\pi)^2}e^{-i{\bm q_{\perp}\cdot\bm x}} \delta\tilde \bE({\bm q}_{\perp})e^{-\frac{|{\bm q}_{\perp}|^2}{2|{\bm B}_{\rm ex}|}}
\\\nonumber
&&=\frac{|{\bm B}_{\rm ex}|}{2\pi}
\int^{\infty}_{-\infty} d^2x'_\perp 
\delta\bE(\bx'_{\perp})e^{-\frac{|{\bm B}_{\rm ex}|}{2}
|\bx_\perp-\bx_\perp'|^2}
\\\nonumber
&&=\frac{|{\bm B}_{\rm ex}|}{2\pi}\int^{\infty}_{0}dr'\int^{2\pi}_{0}d\theta'r'\delta\bE(\bx_\perp')
\nnb
&&\times\exp\Big[-\frac{|{\bm B}_{\rm ex}|}{2}(r'^2+r^2
-2x r'\cos\theta'-2y r'\sin\theta')\Big].
\nn
\end{eqnarray}
For a rotationally symmetric field 
$
\delta\bE(r')$, 
it further reduces to
\begin{eqnarray}
&&\hspace{-0.5cm}
\int \frac{d^2{\bm q}_{\perp}}{(2\pi)^2}e^{-i{\bm q_{\perp}\cdot\bm x}} \delta\tilde \bE({\bm q}_{\perp})e^{-\frac{|{\bm q}_{\perp}|^2}{2|{\bm B}_{\rm ex}|}}
\\\nonumber
&&=|{\bm B}_{\rm ex}|\int^{\infty}_{0}dr'r'\delta\bE(r')e^{-\frac{|{\bm B}_{\rm ex}|}{2}(r'^2+r^2)}I_{0}(|{\bm B}_{\rm ex}|r r'),
\end{eqnarray}
where $I_{n}(z)$ is the modified Bessel function of the first kind.
Taking $\delta\bE(r')=E_0e^{-r'^2/R_{E}^2}$, 
one can analytically carry out the integral
\begin{eqnarray}
	\int \frac{d^4{q}}{(2\pi)^4}e^{iq\cdot x-\frac{|{\bm q}_{\perp}|^2}{2|{\bm B}_{\rm ex}|}} \delta\tilde \bE(q)
	= \frac{E_0e^{-\frac{|{\bm B}_{\rm ex}|r^2}{2+|{\bm B}_{\rm ex}|R_{E}^2}}}{(2\pi)^2\left(1+\frac{2}{|{\bm B}_{\rm ex}|R_{E}^2}\right)}
    ,
\end{eqnarray}
and arrive at the final result 
shown in Eq.~(\ref{eq:current-density-result}):
\begin{eqnarray}
	\delta {\bm j}(\bx_\perp)
    =  \hat{\bB}_{\rm ex}
    \frac{\tau_5|\bB_{\rm ex}|^2E_0}{4\pi^4\chi}
    e^{-\frac{r^2}{R_E^2}}
    \left(1-\frac{e^{\frac{2r^2}{R_E^2(2+\bar{B})}}}{ 1+ 2/\bar B}\right).
\end{eqnarray}

\section{Weak-field limit}
Here, we address the weak magnetic field limit. 
Taking this limit involves a transition between the two series representations in terms of the Landau-level summation 
and the perturbative summation, shown in Fig.~\ref{fig:triangle2}. 
The transition between the two different series representations is not achieved unless one 
performs the infinite summation first to construct 
the complete functional form and then expands it 
in the other series representation. 

The central idea is to introduce the proper-time integral for the denominator in Eq.~\eqref{eq:w_PS}, and to express the sum over the Landau levels in $w_{\rm PS}$ by the generating function~\eqref{eq:Gen}. 
Applying an identity 
$1/(A - i\epsilon) 
= i \int_0^\infty d\tau
\exp\{ -i \tau (A - i \epsilon) \}$
and then taking $t=\exp(-i b \tau x-\epsilon\tau)$ and $s=\exp[- i b \tau(1-x)-\epsilon\tau]$, 
we find that 
\begin{equation}
\label{eq:S2_epsilon}
w_{\rm PS}
 = - i\,\int_0^1 d x\,\int_0^\infty d\tau\, e^{-i \tau [1-\lambda x(1-x)-i\epsilon]+\xi_\perp F(x,\tau)},
\end{equation}
where
\begin{eqnarray} 
F(x,\tau) = -\frac{[1-e^{- i b \tau x}][1-e^{-i b \tau(1-x)}]}{1-e^{- i b  \tau-\epsilon \tau}}  .
\end{eqnarray}
The infinitesimal $\epsilon$ ensures the convergence of the $\tau$-integral and avoids the essential singularities of $F$. 
Expanding $F$ with respect to $b \ll 1$,
one can systematically organize a perturbative series 
\begin{eqnarray}
    w_{\rm PS}  
    = - \sum_{n=0}^\infty  \Big( \frac{ |eB|^2}{ m^{4} } \Big)^n  
   \int_0^1 dx\, c_{n} (x),
\end{eqnarray}
where $c_n$ is the $x$-dependent coefficient, e.g.,
\begin{subequations}
 \label{eq:}
 \begin{eqnarray} 
c_0 \=  \frac{m^2}{m^2-x(1-x)q^2 - i\epsilon} 
\label{eq:PS-LO}
,
\\
c_1 \= 
\frac{2 m^6|\bq_\perp|^2 \, x^2(1-x  )^2}{[m^2-x(1-x)q^2- i\epsilon]^4}.
\end{eqnarray}
\end{subequations}
The remaining integrals can also be expressed with elementary functions. 
While $0<q^2<4m^2$ is assumed for the convergence of the $\tau$  integral, this condition can be  relaxed with analytic continuation after the integrals are performed. 
The leading-order result (\ref{eq:PS-LO}) agrees with Eq.~(20) in Ref.~\cite{Adler:1969gk} when 
one of the photon momenta is vanishing (for a constant magnetic field) as well as those in Refs.~\cite{Schwinger:1951nm, Copinger:2018ftr}.  
Note that the AWI in Eq.~(14) of Ref.~\cite{Copinger:2018ftr} is augmented with the nonlinear terms in a constant electric field, but is still linear in a constant magnetic field. 
The absence of nonlinear terms in a magnetic field from $n\geq 1$ may be due to the absence of spatial modulation.

\bibliographystyle{apsrev4-2}
\bibliography{bib}

\end{document}